# Interrelationship Between Protein Electrostatics and Evolution in HCV and HIV Replicative Proteins

Christopher M. Frenz
Frenzyme Research
cfrenz@gmail.com

**Abstract -** *Protein electrostatics have been demonstrated to play a vital role in protein functionality, with many functionally important amino acid residues exhibiting an electrostatic state that is altered from that of a normal amino acid residue. Residues with altered electrostatic states can be identified by the presence of a pKa value that is perturbed by 2 or more pK units, and such residues have been demonstrated to play critical roles in catalysis, ligand binding, and protein stability. Within the HCV helicase and polymerase, as well as the HIV reverse transcriptase, highly conserved regions were demonstrated to possess a greater number and magnitude of perturbations than lesser conserved regions, suggesting that there is an interrelationship present between protein electrostatics and evolution.*

## 1 Introduction

Rapid advances in the field of structural genomics have resulted in an increased demand for the development of techniques that can provide insights into the functions of proteins based on structural information. Protein electrostatic states, facilitated by interactions between ionizable and polarizable groups, have been demonstrated to contribute to protein catalysis, ligand binding, protein stability, and protein solubility, as well as other aspects of protein function. The Poisson-Boltzmann equation of continuum electrostatics provides a basis for the calculation of electrostatic interactions within proteins, and the finite difference forms of the equation, employed by program packages such as UHBD, APBS, DELPHI, and MEAD, allow for an atomic level of detail[1-4].

Recently, the electrostatic analysis of protein structures has been utilized to determine the protonation states and pKa values of ionizable amino acid residues within protein structures. Calculations of amino acid protonation states across a range of pH values have been used to construct theoretical microscopic titration curves for amino acid residues within the proteins triose phosphate isomerase, aldose reductase, and phosphomannose isomerase. For each of these proteins it was observed that residues within the active sites exhibited perturbed titration curves, in which the residues were partially protonated over a wide pH range. These perturbed titration curves result in a perturbed pKa value for the residues, which, in turn, allows them to more readily contribute to an acid-base catalytic reaction. The proton transfer involved in such reactions can occur most readily when the pKas of the donors and acceptors approach the reaction pH and as such electrostatic interactions are able to alter the residue pKas so that they approach the physiological pH of 7[5]. Residues with pKa perturbations have been demonstrated to play a role in protein stability, such as in pyrrolydone carboxyl peptidases, where E192 was demonstrated to have a pKa value of ≥7.3. This pKa perturbation allows E192 to remain protonated and form a stability contributing strong H-bond with the residue P162[6]. Moreover, residues with shifted pKa values have been observed in residues that contribute to the formation of protein-protein complexes, where 80% of acidic residues within complexes demonstrated a negative shift that helped to preserve their ionized state[7]. Furthermore, studies of the HCV NS3 helicase have suggested that pKa perturbations also exist in residues that contribute to ligand binding, since within the RNA binding cleft of the protein E493 was observed to have a pKa value that was shifted in the positive direction, enabling the residue to remain neutrally charged and allow for improved coordination with DNA[8].

Recent work involving the pKa calculation of 36,192 ionizable residues located in 490 proteins revealed that about 35% of residues possessed a pKa value that was shifted by three or more pK units, and that these perturbations are more likely to occur in buried residues. The study also suggested that buried ionized residues were more likely to be conserved than ionized surface residues[9]. Comparisons between sequence entropy calculations and residue packing in a set of 130 proteins has suggested that tightly packed regions of protein structure are well conserved and that less densely packed regions exhibit lower levels of residue conservation. The rationale behind this is that a protein structures ability to accommodate mutation is restrained by the available space within the structure to accommodate the mutation, and hence more densely packed structures have less space and hence less ability to accommodate a mutation[10].

While the relationship between pKa perturbations and residue conservation is of great interest to the study of all proteins, there are direct clinical applications for such a relationship in the development of antiviral and antimicrobial drug treatments. Both RNA viruses and retroviruses have been demonstrated to rapidly develop antiviral mutations when exposed to drug therapy, since these types of viruses lack the ability to proofread during nucleic acid synthesis. It is hypothesized that within Hepatitis C Virus (HCV) and HIV that there will exist a relationship between residue conservation and pKa perturbation, and that many of these perturbed residues will be of functional significance. This study will examine this relationship within the HCV NS3 helicase and the NS5B polymerase, as well as the HIV reverse transcriptase (RT). Within the HIV RT the study will further examine the relationship between pKa perturbations and antiviral mutations, in order to better examine the role of electrostatics in systems of evolutionary selective pressure.

# 2 Methods

## 2.1 Structure Selection

Structures were selected for each of the proteins examined in this work based on the criteria that one structure exists in the native conformation of the protein and is not bound to any non-native ligands (e.g. an antiviral drug). If available, a structure bound to each native ligands, however, was selected since it was determined that perturbations are only induced in certain residues in the presence of bound ligand[8]. In cases where multiple crystal structures of the unbound or ligand bound states were available only one was selected for analysis. This determination was made because of the computationally intensive nature of the pKa calculation, and the similarity of results found between different crystal structures representative of the same conformation that were computed during trial runs.

**HCV helicase**

The HCV helicase structures used in this study include the 8OHM structure, which is representative of the non-ligand bound form of the enzyme. The 1A1V structure contains a short segment of bound deoxyuracil, which was modified to deoxycytosine to facilitate recognition by the H++ program. Both the 8OHM and 1A1V structures consist of a truncated form of the protein that contains only the helicase region, but lacks the protease region. The third helicase structure considered was the 1CU1 structure which consists of the full length form of the enzyme (helicase and protease regions). This structure is a dimer, but calculations were performed on an individual subunit.

**HCV polymerase**

The HCV polymerase structures utilized in the study include the 1NB4 structure, which is representative of the polymerase in the absence of ligands. The 1NB6 structure is a UTP bound form of the polymerase and the 1NB7 structure consists of the polymerase bound to a short RNA strand. The selenomethionine residues in each structure were modified to methionine to facilitate recognition by the H++ program.

**HIV reverse transcriptase**

The HIV reverse transcriptase structures analyzed included the 1RTJ structure, which is representative of a non-ligand bound form of the enzyme. The 2HMI structure was also analyzed, which consists of the polymerase bound to a double stranded DNA segment. In both cases the structures are dimers consisting of p66 and p50 subunits and electrostatic calculations were performed on the dimers since the dimers are shown to be the biologically active form of the enzyme.

## 2.2 Calculation of pKas

pKa values were calculated using the H++ Web Server, available at http://biophysics.cs.vt.edu/H++ [11]. This Web server begins the pKa prediction by adding missing atoms and assigning partial charges to the uploaded protien structure using the PROTONATE and LEAP modules of the AMBER molecular modelling system and the parm99 force field[12]. The positions of these added protons are then optimized using 100 steps of conjugate gradient descent minimization and 500 steps of Molecular Dynamics simulation at 300K. Protein electrostatic calculations are then performed using the Poisson-Boltzman equation in the program package MEAD, and used to compute the free energies of the protonation microstates[2]. Titration curves and pKa values are then determined using the clustering approach described by Gilson[13].

## 2.3 Calculation of ΔpKa

ΔpKas were calculated by taking the calculated pKa of a given residue and subtracting from it the typical pKa of a residue of that type (Asp=3.65, Glu=4.25, His=6.0, Tyr=10.06, Lys 10.52, and Arg=12.48). The absolute value of the result of this subtraction was then taken. These steps were repeated for each residue of a given enzyme type and the largest ΔpKa value of a given residue across all structures of that enzyme type was selected as the maximum ΔpKa for that residue. The rationale behind this is that different perturbation states are often observed between ligand bound and non-ligand bound forms of the enzyme. Thus taking the maximum pKa observed in any structure is most likely to ensure that the residue is considered in its most electrostatically significant state. For the HIV reverse transcriptase, the p50 subunit is a truncated form of the p66 subunit. Thus the maximum ΔpKa was determined based on the maximum value achieved in any subunit in either of the structures examined. The rationale behind this is that residues may not be of equal significance in each subunit conformation and this method ensures that the most electrostatically significant form of the residue is considered during later residue conservation comparisons.

## 2.4 Determination of Sequence Entropy

Sequence entropy is a measure of relative conservation that takes into account both the frequency and variety of amino acids observed in each position and entropy calculations were performed using the Entropy-ONE Web application available from the HCV Sequence Database Website (http://hcv.lanl.gov/content/hcv-db/ENTROPY/entropy_one.html). Aligned sequences for the HCV NS3 helicase (156 sequence) and HCV NS5B polymerase (149 sequences) were obtained from the database and used to calculate the entropy score associated with each residue position. A set of 400 HIV-1 sequences was obtained from the HIV Sequence Database (http://www.hiv.lanl.gov/content/hiv-db/mainpage.html) and used to compute the associated entropy scores.

## 2.5 Statistical Analysis

All plots and statistical analyses were performed in Graph-Pad Prism version 4.01.

# 3 Results

## 3.1 Residue Perturbations

Many of the perturbed residues in each of the three proteins have been previously characterized as being crucial to the function of the protein. For example, within the HCV helicase the pKa of E490 was shifted upwards by 4.8 pK units, which would mean that the residue would be neutrally rather than negatively charged at physiological pH. This pertubation has been demonstrated to be essential for the coordination of DNA with E493 within the HCV helicase, since it makes residue interaction with negatively charged DNA possible[8]. Likewise in the HCV polymerase D318 and D319 exhibit downward perturbations by 3.1 and 4.9 pK units, respectively, which ensure that the residues always remained within their negatively charged state. Since these residues are involved with the coordination of Mg2+ ions, these pertubations are consistant with their functionality[14]. Within the HIV RT, similar perturbations are observed in residues D185 and D186, which displayed upward pKa shifts of up to 8.8 and 12.6 pK units respectively. These residues have been demonstrated to be involved in the coordination of DNA/RNA and these perturbations would mean that the aspartate residues would be neutrally charged at physiological pH and hence better able to coordinate with DNA[15]. These findings are thereby consistant with previous studies that indicate a correlation between residue functionality and residue perturbation.

## 3.2 Relationship Between Perturbation and Conservation

For each of the proteins invesitgated in this study, plots of ΔpKa versus residue conservation demonstrate that the greatest numbers of perturbed residues are found in low entropy (highly conserved) regions of the proteins. Furthermore, the better conserved residues also exhibited peturbations of greater magnitude than lesser conserved residues (Figure 1).

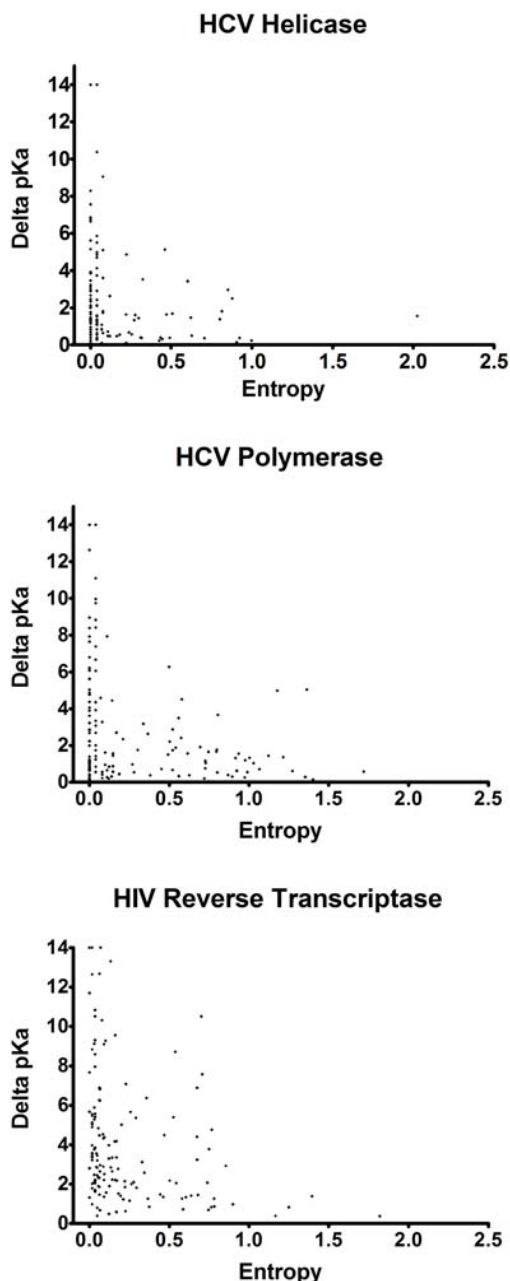

**Fig. 1.** Plot of the magnitudes of residue perturbations versus the entropy scores associated with the reside positions.

The significance of these observations was assessed by dividing the residues into three categories, based on their respective entropy scores. Highly conserved residues were residues with an entropy score of less than 0.1, whereas moderately conserved residues were residues with an entropy score greater than 0.1 but less than 0.4. All residues with an entropy score greater than 0.4 were considered poorly conserved. These groups were created by examining the distribution of residues in Figure 1and dividing the residues into three approximately equal groupings. The fraction of peturbed residues present in each group was then computed (Figure 2) and the Kruskal-Wallis test used to determine that the three groups are statistically significant (P<0.0001).

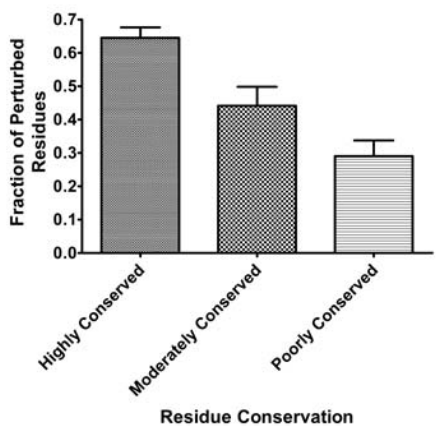

**Fig. 2.** Fraction of perturbed residues present at each level of residue conservation.

The average maximum perturbation magnitude was then computed for each group (Figure 3) and a one-way ANOVA used to determine that each group is statistically significant (P<0.0001). These results are indicative that there is a relationship between electrostatic perturbation and residue conservation, which is likely due to the functional significance of residue perturbation.

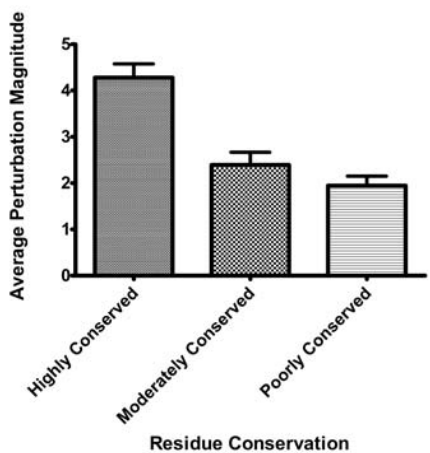

**Fig. 3.** Average magnitude of residue perturbation present at each level of residue conservation.

## 3.3 Effects of Drug Resistance

The relationship between drug resistance and protein electrostatics was examined by compiling a list of all known mutations in ionizable residues that confer drug resistance and examining the perturbation states of these residues (Table 1). Currently out of the 57 residue positions whose mutation has been shown to lead to resistance, there are 11 such mutations in ionizable residues (HIV Sequence Database), and only three of these mutations occur in residues without perturbed pKas. Given the relationship between residue perturbation and functionality these findings are suggestive that the evolution of resistance mutations may alter protein functionality (e.g. ligand binding) via the selected mutation of functionally significant residues.

**Table 1.** Positions of ionizable residues and their associated perturbations and entropy scores

| *Residue* | *Resistance Type* | *ΔpKa* | *Entropy* | *Reference* |
|---|---|---|---|---|
| E44 | 3TC | 2.09 | 0.031 | Montes, 2002 |
| K65 | 3TC | 10.32 | 0.079 | Bazmi, 2000 |
| D67 | AZT | 9.56 | 0.161 | Imamichi, 2000 |
| K70 | AZT | 1.23 | 0.213 | Kellam, 1992 |
| E89 | PFA | 5.41 | 0.035 | Tachedjian, 1995 |
| K101 | AZT + atevirdine | 10.52 | 0.035 | Demeter, 1998 |
| K103 | atevirdine | 2.21 | 0.091 | Demeter, 1998 |
| E138 | TSAO | 2.23 | 0.143 | Van Laethem, 2000 |
| R211 | AZT + 3TC | 0.385 | 1.166 | Kemp, 1998 |
| K219 | AZT | 2.84 | 0.134 | Kellam, 1992 |
| K238 | AZT + atevirdine | 0.251 | 0.251 | Demeter, 1998 |

# 4   Discussion

The results of this study demonstrate that for the HIV reverse transcriptase and the HCV helicase and polymerase there is an association between residue perturbation and residue conservation. This association is supported in that perturbations have been demonstrated in numerous studies to correlate to functionally important elements of a given protein and previous findings that indicate that functionally important residues tend to exhibit a high degree of conservation. These findings also further suggest that protein physics and evolution are highly intertwined. The intramolecular electrostatic interactions of the protein are what account for the presence of these perturbations, and contribute to protein functionality, yet evolution plays a role in the maintenance of these perturbations since the conditions of tight residue packing that enable such interactions are entropicly unfavorable, and would not otherwise be maintained without the presence of some biological advantage.

This inter-relationship between physics and evolution also raises some interesting possibilities when one considers the apparently outlying points on the plots shown in figure 1. Within the plots there are several points that exhibit significant electrostatic perturbations (>2 pK units) but are not well-conserved residues. While there is the possibility of these being the result of calculation errors since pKa calculation software is not always accurate[16], there is also the potential for such residues to be agents of evolutionary change. Since residue perturbations are demonstrated to have functional impact on a protein structure it is a likely possibility that the residue variant present in the crystal structure is having some form of functional impact on the protein structure in question, where functional impact could include possibilities such as a change in catalytic rate, binding affinity, or stability. If they have a positive functional effect it is likely that such residues may become conserved over evolutionary time and if they have a deleterious effect it is likely they will disappear over the course of evolutionary time. It would thus make an interesting avenue of future study to examine the fate of such residues over successive generations of any rapidly replicating organism.

While the above conjecture may explain a possible general mechanism by which the functionality of a given protein can enhance/adapt its suitability to its environment over evolutionary time, the drug resistance data provides a glimpse at evolution in the face of selective pressure. Drug resistance mutations generally seek to reduce the protein's capacity for binding the drug and in doing so often reduce the protein's ability to bind its native ligand as well resulting in reduced enzyme fidelity and/or activity. Since 8 out of 11 drug resistance mutations of ionizable residues occurred in perturbed positions, it suggests that the evolution of drug resistance mutations targets residues of functional significance. This is supported by findings, which indicate that several of the drug resistance residue positions have been characterized as being functionally significant. The mutation of E44 to an aspartate has been demonstrated to cause a decrease in viral load, indicating that the enzyme functionality is reduced by the mutation of this position[17]. Furthermore, K65 has been demonstrated to have a role in nucleotide binding specificity[18] and K103 has been demonstrated to be critical to the transition of the unliganded nucleotide binding pocket from its closed to its open state[19]. It would be of great interest if future studies could be utilized to determine if mutation of functionally significant residues is limited to the development of drug resistance or is a characteristic common amongst multiple cases of evolution under selective pressure.

Moreover, the conservation of perturbed residues observed in the absence of selective pressure indicates that these perturbations and their associated functionality are also likely conserved, and this may likely form the basis of the some of the common functional motifs that are observed. For example, in the DEXH ATP binding motif of the HCV helicase, which is analogous to the DEAD box binding motif and has been previously demonstrated to have functionaly significant perturbations[8], component residues have strong electrostatic interactions with each other that contribute to these perturbations, according to the residue interaction data generated by the H++ program. It was also observed throughout the data set, that many of the residues which strongly interacted with perturbed residues were also highly conserved, suggesting that the functionality provided by many residues may be to help provide a proper electrostatic environment for the maintanence of functionaly significant perturbations. Whether such electrostatic perturbation and interaction motifs are conserved over evolutonary time remains to be shown.

In all, the findings of this study are indicative that an intricate relationship exists between protein physics and evolution, but that future studies are needed to better define this relationship and the mechanisms under which it operates. Once understood, however, knowledge of this relationship could be used to not only enhance our understanding of protein evolution, but could also likely provide the basis for powerful tools for protein engineering and *de novo* protein design.